# Local vaccination and systemic tumor suppression via irradiation and manganese adjuvant in mice

Chunyang Lu[1#], Jing Qian[2#], Jianfeng Lv[1], Jintao Han[1], Xiaoyi Sun[1], Junyi Chen[1], Siwei Ding[2], Zhusong Mei[1], Yulan Liang[1], Yuqi Ma[1], Ye Zhao[2], Chen Lin[1], Yanying Zhao[1], Yixing Geng[1], Wenjun Ma[1], Yugang Wang[1], Xueqing Yan[1] and Gen Yang[1, 3*]

1 State Key Laboratory of Nuclear Physics and Technology, School of Physics & CAPT, Peking University, Beijing 100871, China

2 Teaching and Research Section of Nuclear Medicine, School of Basic Medical Sciences, Anhui Medical University, Hefei 230032, China

3 Wenzhou Institute, University of Chinese Academy of Sciences, Wenzhou 325001, China

#These authors contribute equally.

* Correspondence:

Corresponding Author: Dr. Gen Yang (gen.yang@pku.edu.cn)

**Abstract**

Presently 4T-1 luc cells were irradiated with proton under ultra-high dose rate FLASH or with gamma-ray with conventional dose rate, and then subcutaneous vaccination with or without Mn immuno-enhancing adjuvant into the mice for three times. One week later, we injected untreated 4T-1 luc cells on the other side of the vaccinated mice, and found that the untreated 4T-1 luc cells injected later nearly totally did not grow tumor (1/17) while controls without previous vaccination all grow tumors (18/18). The result is very interesting and the findings may help to explore *in situ* tumor vaccination as well as new combined radiotherapy strategies to effectively ablate primary and disseminated tumors. To our limited knowledge, this is the first paper reporting the high efficiency induction of systemic vaccination suppressing the metastasized/disseminated tumor progression.

**Introduction**

Radiotherapy (RT) is a conventional therapy for the treatment of local tumors or isolated metastases, it exerts cell killing effect by inducing DNA damage.[1] In addition to local anti-tumor effects, RT also has a systemic inhibitory effect on non-irradiated tumors. This phenomenon is called "abscopal effect". The emergence of abscopal effect solved the existing dilemma of radiotherapy, that is, it can only treat local tumors. At present, the specific mechanism of abscopal effect is still unclear, but studies have shown it may be related to the immune response,[2] including the improvement of T cell activation/ initiation by increasing the local availability of tumor-associated antigen (TAA) or promoting the release of immune-stimulating cytokines.[3] And new evidence from pre-clinical and clinical studies also proved that the combination of radiotherapy and immunotherapy can enhance the body's anti-tumor response more effectively than either therapy alone.[4,5] More and more studies have shown that for cancer patients, especially late-stage cancer patients, the combined treatment method (radiotherapy and immunotherapy) seems to be safe and can significantly improve the response of tumor treatment.[4]

The cGAS-STING pathway, which recognizes DNA in cells, is essential for tumor immune surveillance and immunotherapy. Jiang *et al.* conducted in-depth and systematic studies on physiological functions of STING protein, and found the immunomodulatory function of manganese ions. Manganese ions played the dual functions of "alarm element" and "agonist" in the cGAS-STING pathway. Furthermore, it has broad prospects in anti-infection, anti-tumor and immune adjuvant applications in the future.[6]

As metastasis is responsible for as much as 90% of cancer associated mortality, [8-9] *in situ* vaccination as well as efficiently induction of systemic tumor suppression is the key to tumor treatment. In this study, we injected 4T-1 luc cells (after FLASH or γ ray irradiation) and/or Mn immuno-enhancing adjuvant into the mice for three times. One week later, we injected untreated 4T-1 luc cells on the other side of the mice, and found that the 4T-1 luc cells injected later did not grow tumor while controls without previous vaccination all grow tumors. The result is very interesting and the findings may help to explore *in situ* tumor vaccination as well as new combined radiotherapy strategies to effectively ablate primary and disseminated tumors.

## Materials and methods

### Cell culture

4T1-luc cells were cultured in RPMI medium (Hyclone) supplemented with 10% fetal bovine serum (FBS, Bai Ling Biotechnology) and 1% penicillin-streptomycin (P/S) (Hyclone). All cells were cultured in a humidified incubator at 37°C and 5% $CO_2$ (Sanyo).

### Preparation of Mn adjuvant

As reported previously [6], 50 μL $Na_3PO_4$ (Klamar) were added to 850 μL physiological saline, and then add 100 μL $MnCl_2$ (0.2 M, Solarbio), overnight aging to make colloidal manganese $Mn_2OHPO_4$.

### FLASH irradiation

The FLASH radiation experiment was performed using Compact LAser Plasma Accelerator (CLAPA) system in Peking University as reported previously.[7] In order to facilitate the irradiation, 3.5 μm Mylar membrane was glued to the bottom of the cell-culture hoop in advance, and were pretreated with RPMI medium/1% Geltrex (Gibco) for 24 h. Cells were seeded 8 h before irradiation. 4T1-luc cells were to 150,000 cells/mL, and 800 μL cell suspension were added to each cell-culture hoop, before starting the irradiation, the medium in cell-culture hoop was aspirated, and the cell-culture hoop was screwed on the hoop holder and mounted vertically. Then proton beams with ultra-high dose rate irradiated the cells, after that, cells were digested and injected into mice's left breast pad (day 0). On day 1, cells undergo the same irradiation operation but different dose, and were digested and injected into the left tail of the mice. On day 8, cells undergo the same irradiation operation but different dose, and were digested and injected into the left tail of the mice. The doses of three irradiations were 20 Gy (day 0), 25 Gy (day 1), 105 Gy (day 8). On day 15, 100,000 untreated 4T-1 luc cells were injected into mice's left breast pad. On day 22 and day 29, the substrate was injected into the abdominal cavity and observed by imaging.

**γ ray irradiation**

Cells were seeded into 24-well plates 8 h before irradiation. 4T1-luc cells were to 150,000 cells/mL, and 800 μL cell suspension were added to each well. And then were irradiated with γ ray. After that, cells were digested and injected into mice's left breast pad (day 0). On day 1, cells undergo the same irradiation operation but different dose, and were digested and injected into the left tail of the mice. On day 8, cells undergo the same irradiation operation but different dose, and were digested and injected into the left tail of the mice. The doses of three irradiations were 20 Gy (day 0), 25 Gy (day 1), 105 Gy (day 8). On day 15, 120,000 untreated 4T-1 luc cells were injected into mice's left breast pad. On day 22 and day 29, the substrate was injected into the abdominal cavity and observed by imaging.

**Live imaging to observe tumor growth**

Tumor growth was observed by the small animal live imaging system on day 22. Before the observation, the chest skin of each mice was fully depilated with depilatory cream. Then mice were weighed, and the substrate (D-Luciferin potassium salt, 10 mg/mL, D&B biotechnology) was injected into the abdominal cavity of the mice, according to 20 g mice corresponding to 2 mL substrate, calculate the amount of substrate required for each mouse. After the substrate was injected for 25 minutes, mice were anesthetized with isoflurane inhalation for 3 minutes, and were put into the small animal live imaging system for imaging (Caliper life sciences, IVIS Lumina III), use Spectrum Living Image 4.0 software for analysis. Perform the imaging again on day 29, the mice were immediately sacrificed after imaging, and the tumor was dissected out. All animal experiments and protocols were approved by the Peking University Institutional Animal Care and Use Committee (Approval ID: Physics-YangG-2).

# Results

## Grouping of mice and implementation of irradiation

The mice were divided into four groups, group I and group II were control groups, group III and group IV were experimental groups. The mice in group I received no treatment. The mice in group II were injected with 9.1 μL Mn adjuvant on the left breast pad on day 0. The mice in group III were injected with irradiated 4T1-luc cells on the left breast pad three times on day 0, day 1 and day 8, as described above. And the mice in group IV were injected with irradiated 4T1-luc cells along with Mn adjuvant on the left breast pad three times on day 0, day 1 and day 8. On day 15, all mice (group I -IV) were injected with 100,000 untreated 4T-1 luc cells on the left breast pad, and tumor growth was observed using the small animal *in-vivo* imaging system and was analyzed by the Spectrum Living Image 4.0 software.

## FLASH irradiation

The imaging results showed that tumors appeared on the left side of the mice in the control groups (group I and group II, Figure 2A and B), where the untreated cells were injected. And tumors appeared on the left side of the experimental groups (group III and group IV, Figure 2C and D) where the irradiated cells were injected, but there was no tumor on the left side where the untreated cells were injected.

## γ-ray irradiation

At the same time, tumors appeared on the left side of the control groups (group Ⅰ and group Ⅱ, Figure 3A and B), where the untreated cells were injected. But only one mouse in group Ⅲ had a tumor formation on the left side, and no tumors appeared on the left sides in the mice of experimental groups (Figure 3C and D). It can be inferred that a series of immune reactions occurred after the irradiated cells were injected into the mice, and inhibited the tumor formation of later injected untreated 4T1 cells, but the results of the FLASH groups and the γ-ray irradiation groups were different, which may require further study.

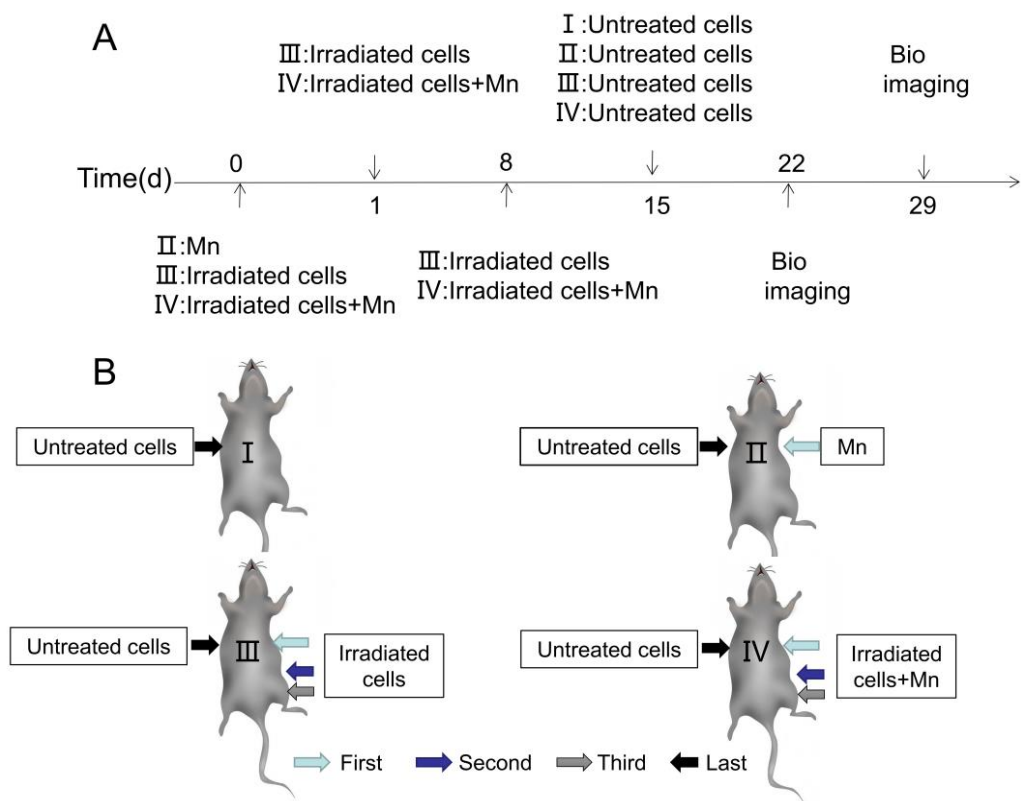

Figure 1 The timeline and the grouping of mice. (A) Timeline for injection

of irradiated/untreated cells and/or adding Mn adjuvant. (B) Detailed grouping of the four groups of mice and specific injection sites of cells/ Mn adjuvant.

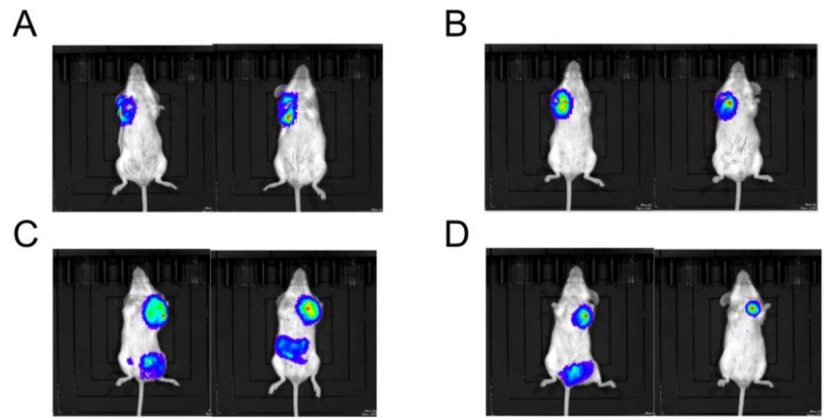

Figure 2 The imaging of mice in FLASH irradiation groups on day 22. (A) The left side of the mice in group Ⅰ developed tumors. (B) The left side of the mice in group Ⅱ developed tumors. (C) Only the left side of the mice in group Ⅲ developed tumors. (D) Only the left side of the mice in group Ⅳ developed tumors.

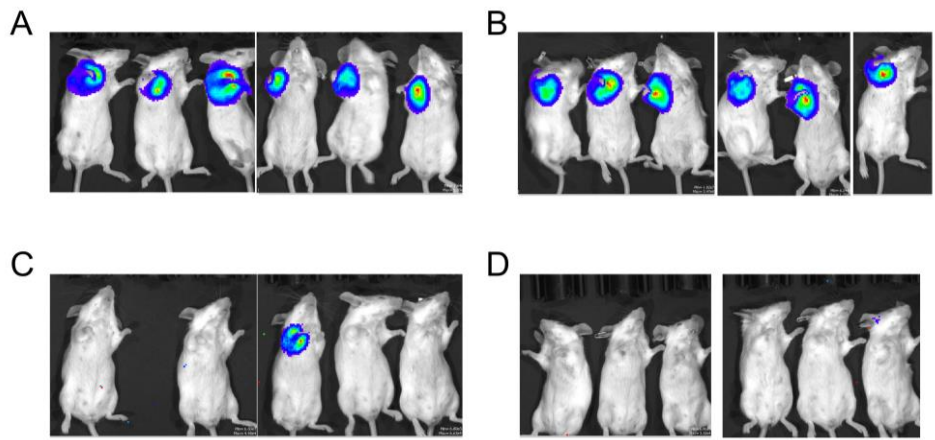

Figure 3 The imaging of mice in γ-ray irradiation groups on day 22. (A)

The left side of the mice in group Ⅰ developed tumors. (B) The left side of the mice in group Ⅱ developed tumors. (C) Only one mouse in group Ⅲ showed tumors on the left side, and none of the other mice showed tumors. (D) None of the mice in group Ⅳ showed tumors.

Table 1 *In vivo* tumor formation ability of FLASH and γ ray irradiation group.

|  | Group Ⅰ |  | Group Ⅱ |  | Group Ⅲ |  | Group Ⅳ |  |
| --- | --- | --- | --- | --- | --- | --- | --- | --- |
|  | Left |  | Left |  | Left | Right | Left | Right |
| FLASH irradiation group | 3/3 |  | 3/3 |  | 0/3 | 3/3 | 0/3 | 3/3 |
| γ ray irradiation group | 6/6 |  | 6/6 |  | 1/5 | 0/5 | 0/6 | 0/6 |

## Discussion

Cancer metastasis is the leading cause of cancer-related deaths.[8,9] Although RT is one of the main treatments for malignant tumors, it has limited efficacy for metastatic disease. Studies have shown RT may cause the abscopal effect that inhibits metastatic lesions, which indicated a potential treatment direction. Using the abscopal effect of RT combined with immunotherapy may provide powerful tools for tumor treatment. However, the absopcal effects are generally rare in very low probability, thus efficiently *in situ* vaccination as well as induction of systemic tumor suppression are the key issues.

In this study, we observed tumor growth by injecting irradiated 4T1-luc cells and untreated 4T1-luc cells on both sides of the mice, and found that the irradiated cells and/or Mn adjuvant can indeed inhibit the tumor growth of untreated 4T1-luc cells, the probability of present study setup are 100%, this is interesting. At the same time, the results showed that FLASH irradiation and γ ray irradiation are different. After FLASH irradiation, the cells formed tumors, but it also inhibited the growth of untreated 4T1-luc cells; while after γ ray irradiation, cells did not form tumors, and it also inhibited the growth of untreated 4T1-luc cells. All the results suggested that irradiation may cause a series of immune responses, and these strong immune responses prompted the mice to systemic suppress the subsequent seeded 4T1-luc cells.

Currently, although the mechanisms are not clear and need urgent study carefully, these results themselves are of great significance. The most interesting aspect is taking good advantages of the phenomenon to generate *in situ* vaccination thus can systemically suppress the metastasized/disseminated tumors. Further understanding of the mechanisms may greatly promote cancer immunotherapy and radio-immunotherapy, thus may impact on the progress of clinic cancer treatment strategies.

## Conflict of Interest

The authors declare that the research was conducted in the absence of any commercial or financial relationships that could be construed as a potential conflict of interest.

## Author Contributions

Gen Yang conceived the research plan. Gen Yang, Chunyang Lu and Jing Qian designed the biological and irradiation experiments. Chunyang Lu and Jing Qian performed the biology experiments, analyzed results, and produced figures. Gen Yang, Zhusong Mei, Chunyang Lu and Wenjun Ma designed and constructed the irradiation setup. Zhusong Mei, Yulan

Liang, Chen Lin, Yanying Zhao and Yixing Geng performed the laser acceleration and cell irradiation experiments. Zhusong Mei and Yulan Liang measured the proton radiation and provided the dose data. Chunyang Lu and Jing Qian wrote the initial draft of the paper. Gen Yang wrote the final draft of the manuscript. All authors commented on the manuscript.

**Funding**

This study was supported by the National Natural Science Foundation of China (11875079, 61631001 and 119210067), the National Grand Instrument Project (2019YFF01014402), NSFC innovation group project (No. 11921006), and the State Key Laboratory of Nuclear Physics and Technology, PKU under Grant No. NPT2020KFY19 and NPT2020KFJ04.

**Acknowledgments**

We thank Jiuqiang Li, Zhuo Pan, Defeng Kong, Shirui Xu, Zhipeng Liu, Ying Gao, Guijun Qi, Yinren Shou, Shiyou Chen and Zhengxuan Cao for the operation of the gamma as well as CLAPA proton irradiations.

**Data Availability Statement**

Data available within the article.